\documentclass[twocolumn]{aastex701} 

\usepackage{multirow}

\usepackage{graphics,epsf}
\usepackage[utf8]{inputenc}
\usepackage{amsmath}                
\usepackage{amsfonts}               
\usepackage{amssymb}                
\usepackage{epsfig}                 
\usepackage{graphicx}               
\usepackage{float}
\usepackage{color}
\usepackage{multirow}               

\hypersetup{
    colorlinks=true,
    linkcolor=red,   
    urlcolor=cyan}

\usepackage[colorinlistoftodos]{todonotes}

\newcommand{\s}{{~\rm s}}

\newcommand{\g}{{~\rm g}}
\newcommand{\G}{{~\rm G}}
\newcommand{\K}{{~\rm K}}

\begin{document}

\title{Jittering jets promote dust formation in core-collapse supernovae}

\author[0000-0003-0375-8987]{Noam Soker}
\affiliation{Department of Physics, Technion, Haifa, 3200003, Israel; soker@physics.technion.ac.il}
\email{soker@physics.technion.ac.il}

\begin{abstract}
I find that the dust morphologies in some core-collapse supernova (CCSN) remnants (CCSNRs) possess jet-shaped morphologies, and propose that the properties of the jets that explode the CCSNe and their interaction with the core and envelope (if it exists) are among the factors that determine the amount of dust formed and its morphology. I find that some of the dust-rich structures in the CCSNRs Cassiopeia A and the Crab Nebula are distributed in point-symmetric morphologies, and that the dust in SN 1987A follows the bipolar morphology of the inner ejecta. Earlier studies attributed these morphologies in CCSNRs to jet shaping within the jittering jets explosion mechanism (JJEM). These dust morphologies suggest, within the framework of the JJEM, that exploding jets enhance dust formation in CCSNRs. This study contributes to the diversity of processes in which CCSN exploding jets are involved and to establishing the JJEM as the primary explosion mechanism of CCSNe.
\end{abstract}

\section{Introduction} 
\label{sec:intro}
One of the largest open questions concerning core-collapse supernovae (CCSNe) is their explosion mechanism. Recent years saw two intensively studied theoretical explosion mechanisms, the delayed neutrino mechanism (e.g., \citealt{Huangetal2025, Imashevaetal2025, Laplaceetal2025, Maltsevetal2025, Maunderetal2025, Morietal2025, Mulleretal2025, Nakamuraetal2025, SykesMuller2025, Janka2025, Orlandoetal20251987A, ParadisoCoughlin2025, Tsunaetal2025, Vinketal2025, WangBurrows2025, Willcoxetal2025, Mukazhanov2025, Raffeltetal2025, Vartanyanetal2025})\footnote{For recent talks on the neutrino-driven mechanism see, e.g., \citealt{Janka2025Padova},  \url{https://www.memsait.it/videomemorie/volume-2-2025/VIDEOMEM_2_2025.46.mp4}, and \url{https://www.youtube.com/watch?v=nRfDPPSmnzI&t=100s}}, and the jittering jets explosion mechanism (JJEM; e.g., \citealt{Bearetal2025Puppis, 
Braudoetal2025, Kumar2025, Shishkinetal2025S147, Soker2025Learning, SokerAkashi2025, WangShishkinSoker2025})\footnote{For a talk on the JJEM see \citealt{Soker2025Padova}: \url{https://www.memsait.it/videomemorie/volume-2-2025/VIDEOMEM_2_2025.47.mp4}}. 
The magnetorotational explosion mechanism, which involves fixed-axis jets (e.g., \citealt{Shibataetal2025}) and applies to rare cases of rapidly rotating pre-collapse cores, attributes most CCSNe to the neutrino-driven mechanism. Therefore, it is not an alternative explosion mechanism for most CCSNe. 

In the JJEM, several to about twenty pairs of opposite jets along different axes that an intermittent accretion disk around the newly born neutron star launches within several seconds, explode the star. The source of the stochastic angular momentum of the intermittent accretion disks is the motion in the convective zones of the pre-collapse core. The convective motion seeds instabilities around the neutron star that amplify the angular momentum fluctuations. The large angular momentum fluctuations of the accreted mass onto the newly born neutron star lead to the formation of the short-lived accretion disks with varying axes. These intermittent accretion disks launch the jittering jets that explode the star.   

The JJEM has several successes in accounting for observations, such as energetic CCSNe (for recent summaries, see \citealt{Soker2024UnivReview, Soker2025Learning}). According to the JJEM, the transition from neutron star to black hole masses is continuous and sparsely populated \citep{Soker2023NSBH}, which explains the mass distribution inferred from merging compact objects (e.g., \citealt{LVK2025}). 

The observable property that best distinguishes the two alternative explosion mechanisms is the morphology of CCSN remnants (CCSNRs; e.g., \citealt{Soker2024UnivReview, Soker2025Learning}). The JJEM predicts that many, but not all, CCSNRs possess point-symmetric morphologies, whereas the neutrino-driven mechanism does not account for point-symmetric morphologies. Point-symmetric morphologies are those where there are two or more pairs of opposite structural features that do not share the same axis. The two structural features in a pair are opposite with respect to the center of the CCSNR, although they do not need to be of the same size and shape, nor at the same distance from the center. Opposite structural features include (e.g., \citealt{Bearetal2017}) dense clumps, filaments, low-density bubbles, open bubbles (termed lobes), protrusions (termed ears; see \citealt{GrichenerSoker2017}), and rings. Many earlier studies have explored point-symmetric CCSNR morphologies of gas radiating in the radio, optical, and X-ray regimes (e.g., \citealt{ShishkinKayeSoker2024, Bearetal2025Puppis, Shishkinetal2025S147, Soker2025N132D, Soker2025RCW89}). 
In this \textit{Letter}, I concentrate on the morphology of dust as revealed by IR radiation of CCSNRs.   
CCSNRs and other supernova remnants reveal much information about supernovae ejecta and their interactions, such as ejecta structure (e.g., \citealt{Renetal2018RAA}), ejecta interaction with the ambient gas (e.g., \citealt{Yanetal2020RAA, Luetal2021RAA, Dedikovetal2025NewA}), including jets (e.g., \citealt{YuFang2018RAA}), different emission properties (e.g., \citealt{Yamazakietal2014RAA, Zhangetal2016RAA, Lietal2020RAA, Orlandoetal2021NewA, Luoetal2024RAA, Mwanikietal2025NewA, YuHetal2022NewA}), magnetohydrodynamics (e.g., \citealt{Wuetal2019RAA, Leietal2024RAA}), and the neutron star remnant  (e.g., \citealt{HorvathAllen2011RAA, Wuetal2021RAA}). Relevant to this study is that many supernova remnants have been studied over the years for their dust (e.g., \citealt{Shahbandehetal2023, Shahbandehetal2025, LuXiWeietal2025RAA}). Here, I only focus on morphologies that I attribute to pairs of jets (Section \ref{sec:Dust}). I argue that these suggest that compression by the jets enhances dust formation in CCSN ejecta (Section \ref{sec:Dust}). In Section \ref{sec:Summary}, I summarize this study that further enriches the JJEM.   

\section{Point-symmetry of dust morphologies}
\label{sec:Dust}

I present IR images of three CCSNRs for which earlier studies identified point-symmetrical morphologies or a prominent bipolar morphology. So, the identification of point-symmetric morphologies is not new. The new claim is that dust features exhibit clear jet-shaped morphologies, suggesting that jets play a significant role in dust formation in CCSNe, to the extent that jittering jets determine the amount of dust formation in some CCSNe.  
I note that dust morphologies in some planetary nebulae are bipolar (e.g., \citealt{Kwoketal2008}, their figures 12 and 13; \citealt{ZhangYKwok2009}, their figure 14; \citealt{ZhangYetal2012}), morphologies that are attributed to jets in the planetary nebula community (see \citealt{Kwoketal2026Galax} for a recent review).  

I do not study the classification of the three CCSNRs as jet-driven here, but accept earlier studies that claimed so. In this \textit{Letter}, I refer only to the dust morphology in relation to the jet-driven morphologies of the CCSNRs. 

To identify structural features, we follow the widely used practice of studying planetary nebula morphologies and classifying them through careful visual inspection and qualitative classification (e.g., \citealt{Balick1987, Parkeretal2006, Sahaietal2007, Kwok2024}; for another emerging approach see \citealt{ShishkinMichaelis2026}). Although qualitative, this method enables the identification of morphological features that jets shaped (e.g., \citealt{SahaiTrauger1998}), comparing them with numerical simulations, and by that shedding light on the shaping processes through qualitative comparisons with simulations (e.g., \citealt{GarciaSEguraetal2022, GarciaSEguraetal2025}), as well as comparing planetary nebulae to CCSNRs (e.g., \citealt{Akashietal2018}). 
The visual inspection used to infer jet shaping has been applied to other types of systems, such as B[e] supergiants (e.g., \citealt{Kashi2023}) and luminous blue variable nebulae (e.g., \citealt{Kashi2024}).  
This method for identifying point-symmetric structures has been successfully applied to CCSNRs in the past (e.g., \citealt{Soker2024W44, Soker2024NA1987A}).

\subsection{Point-symmetry of dust in Cassiopeia A}
\label{subsec:CassiopeiaA}
\begin{figure}
\begin{center}
\includegraphics[trim=5.7cm 3.3cm 1.9cm 0.0cm ,clip, scale=0.43]{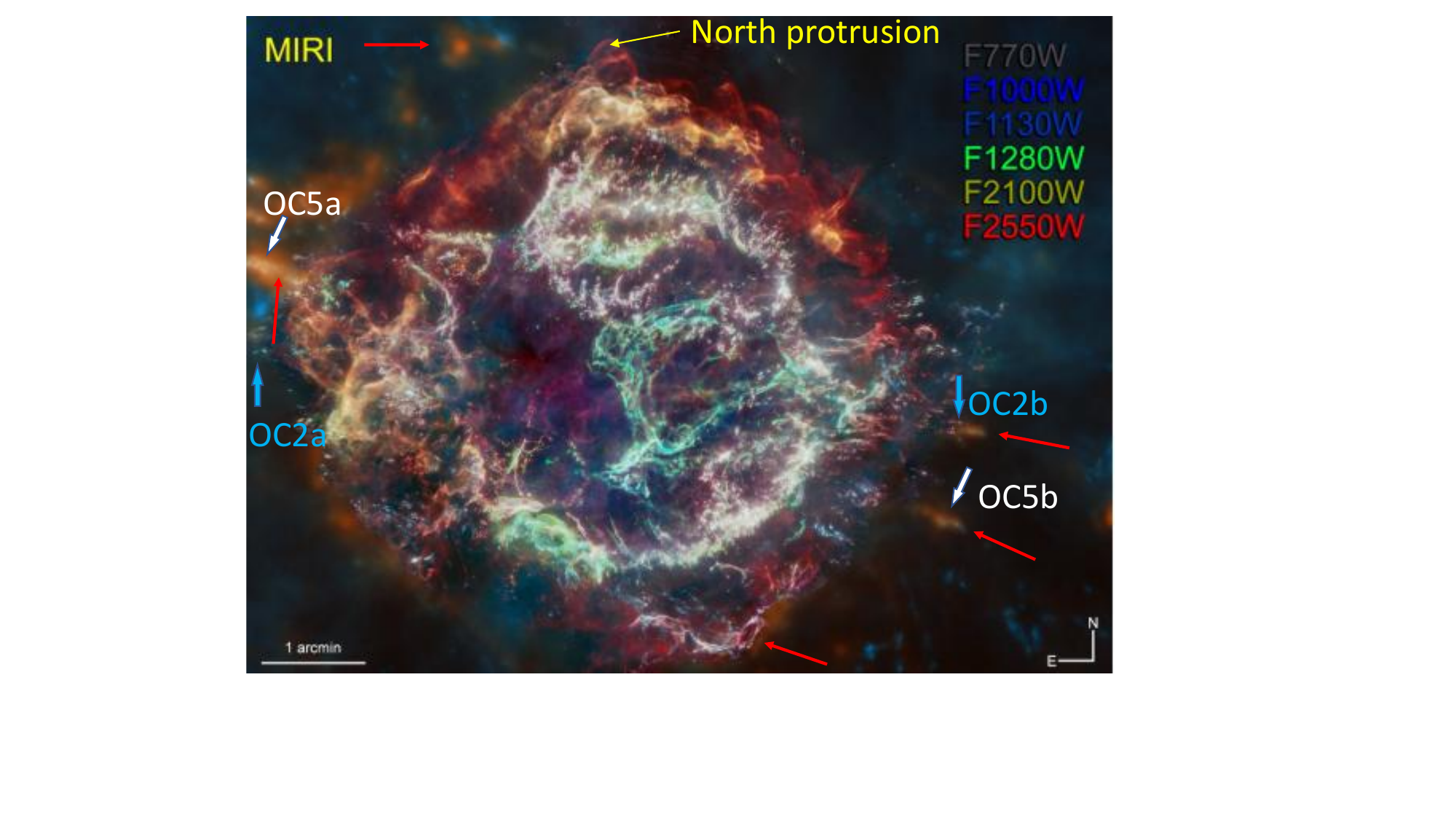}
\includegraphics[trim=5.7cm 3.3cm 1.9cm 0.3cm ,clip, scale=0.43]{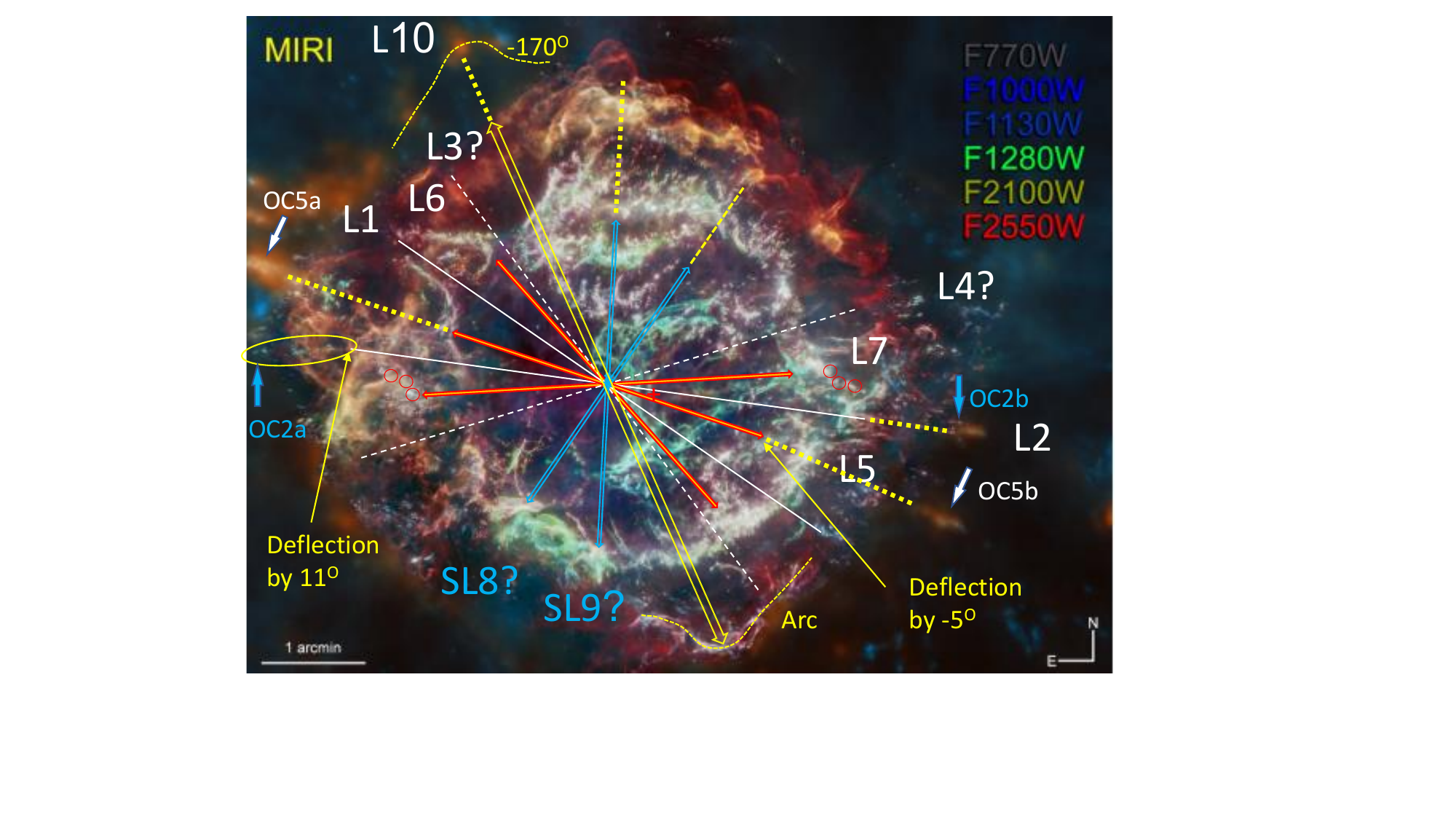}
\end{center}
\caption{A JWST image of Cassiopeia A that \cite{BearSoker2025} adapted from \cite{Milisavljevicetal2024} and added their identifications of the point-symmetrical morphology (lower panel). I added five red arrows on the upper panel, pointing to dust-rich structural features that form pairs within the point-symmetric morphology. 
\cite{BearSoker2025} mark the arc at the south with a dashed-yellow line and copied it to the north of the remnant and rotated it around itself by $-170 ^\circ$ ($170^\circ$ clockwise). Here, I emphasize that these two opposite arcs are dust-rich.       
}
\label{Fig:CassiopeiaJWST} 
\end{figure}

In Figure \ref{Fig:CassiopeiaJWST} I present a JWST image of Cassiopeia A adapted from \cite{Milisavljevicetal2024}, and with additional marks by \cite{BearSoker2025}. Studies have shown that Cassiopeia A contains dust (e.g., \citealt{Barlowetal2010, Arendtetal2014, DeLoozeetal2017, Priestleyetal2022}), but only the high-resolution images by JWST allowed \cite{BearSoker2025} to identify the point-symmetric structure in the IR clearly. The red and orange colors indicate the presence of dust. Most of the dust is located in the north and northeast, which may be related to circumstellar material from pre-explosion evolution, such as in a binary system (e.g., \citealt{Hiraietal2020}). However, the clumps OC2b and OC5b also contain dust. They are part of the rich point-symmetric morphology that \cite{BearSoker2025} identified in Cassiopeia A. The OC5a filament, the counterfeature of OC5b, also appears to contain dust.   

Another possible dusty pair is the two sides of line L10. At the north edge of line L10, there is an orange clump, likely a dusty clump. At the south edge of line L10, there is an orange arc. According to the JJEM, the two opposite arcs marked by \cite{BearSoker2025} with dashed yellow curved lines were compressed by two opposite jets of a pair.   I suggest that the jets compressed the gas, and the high density of the jet-compressed zones facilitated dust formation. 
Note that the general dust distribution in Cassiopeia A has a concentration in the north.

I conclude that some, but not all, of the dusty structures in Cassiopeia A form a point-symmetric structure. I attribute this morphology to jets in the framework of the JJEM.  

\subsection{Point-symmetry of dust in the Crab Nebula}
\label{subsec:Crab}

\begin{figure}
\begin{center}
\includegraphics[trim=0.0cm 0.0cm 0.0cm 0.0cm ,clip, scale=0.49]{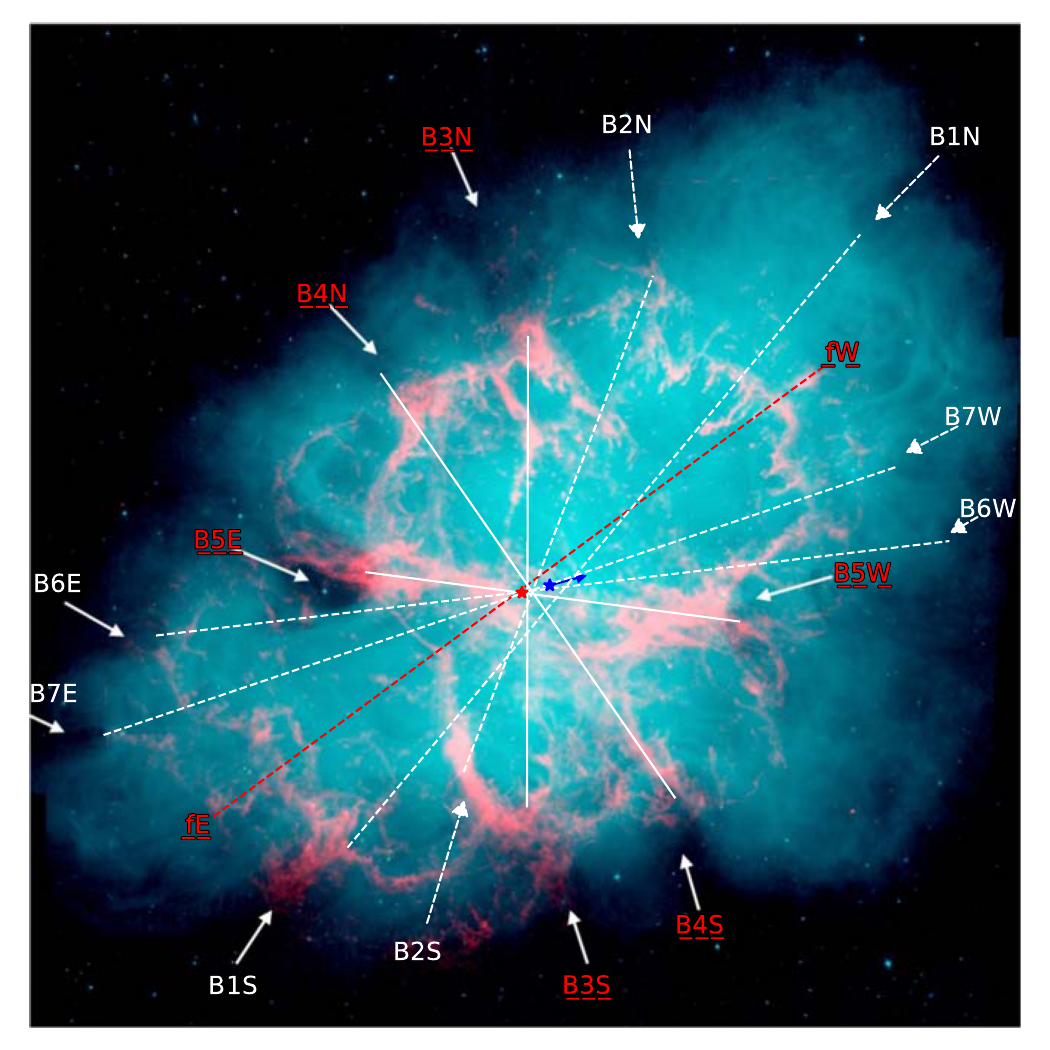}
\caption{An image of the Crab Nebula that \cite{ShishkinSoker2025Crab} adapted from Figure 13 of \cite{Temimetal2024}. Blue represents synchrotron emission, and red represents emission by dust. While \cite{Temimetal2024} identified nine bays with dust filaments, which they marked with solid white arrows, \cite{ShishkinSoker2025Crab} identified five additional bays (dashed-white arrows). \cite{ShishkinSoker2025Crab} identified a rich point-symmetric morphology by connecting the bays identified by \cite{Temimetal2024} (solid white lines) and the bays they identified (dashed white lines), as well as one pair of dust filaments (dashed red line). The blue asterisk marks the location of the pulsar (PSR B0531+21), and the red asterisk is the calculated location of the neutron star at the explosion. The underlined red names refer to the eight dusty filaments that form a dust-rich, point-symmetric morphology. 
Note that the arrows point to the bays, while the dust filaments are behind the bays, closer to the center, in red color. The dust being closer to the center indicates that the enhanced dust formation is not due to ejecta interaction with an ambient gas. I would rather attribute it to gas compression by jets. 
}
\label{fig:CrabDust}
\end{center}
\end{figure}

\cite{ShishkinSoker2025Crab} analyzed an image from \cite{Temimetal2024}, identified a point symmetric morphology in the Crab Nebula, and claimed this is a strong indication that jittering jets exploded the Crab Nebula. I present this image from \cite{Temimetal2024} and with the marks by \cite{ShishkinSoker2025Crab} in Figure \ref{fig:CrabDust}, where blue represents synchrotron emission and red represents emission by dust. I identify four prominent pairs of opposite dust features, whose names are underlined in red. Clearly, some of the dust filaments in the Crab Nebula build a point-symmetric morphology. 

The filaments of dust are mainly in the regions closer to the center than the bays. Therefore, they are unlikely to result from the compression of gas by the collision of the ejecta with an ambient gas. 

I attribute some of the dust morphology to jittering jets and suggest that these jets, which also exploded the star, enhanced dust formation in the Crab Nebula. 

\subsection{Bipolar dust morphology in SN 1987A}
\label{subsec:SN1987A}

Some small clumps of SN 1987A display a possible point-symmetric morphology \citep{Soker2024NA1987A}. However, the main structure of SN 1987A, now a CCSNR, is a bipolar structure, the `Keyhole,' similar in some aspects to some other astrophysical jet-shaped objects, leading to the suggestion of a powerful pair of jets that shaped this structure (e.g., \citealt{Soker2024Keyhole, Soker2024PNSN}). ALMA and JWST observations show that some dust in the ejecta follows the bipolar morphology of the `Keyhole.' 

\begin{figure}
\includegraphics[trim=0.0cm 0.0cm 0.0cm 0.0cm ,clip, scale=0.46]{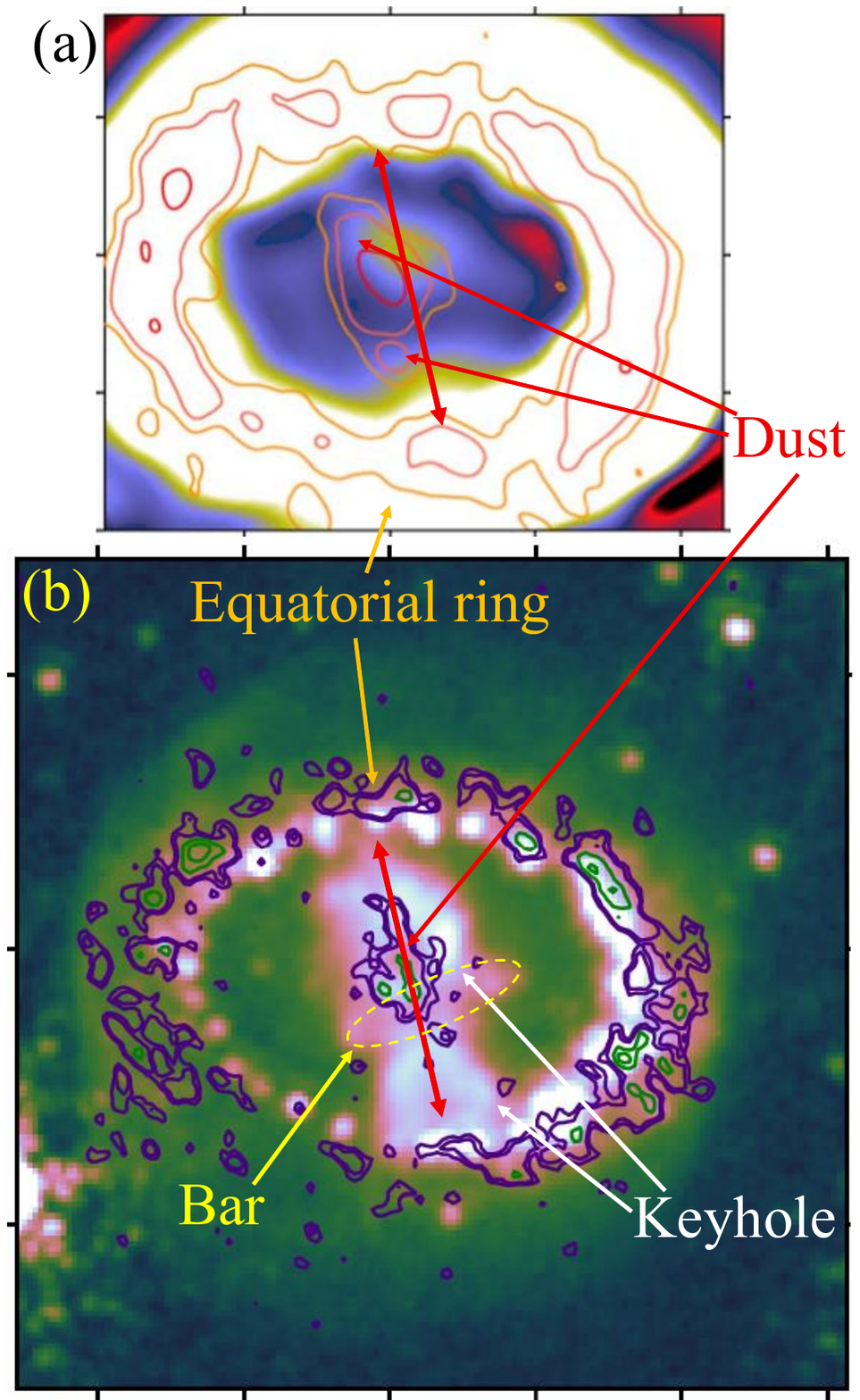}
\caption{ Two images of ALMA and JWST observations, at the same scale, emphasizing the dust elongated structure along the main symmetry axis of SN 1987A, i.e., the axis of the bipolar `Keyhole' structure.  
(a) An image adapted from \cite{Bouchetetal2024}. Colors represent the JWST MIRI F560W image, and the contours show the ALMA 315~GHz image, highlighting cold dust emission. Axes are right ascension, with ticks on the horizontal axis from $5^{\rm h}35^{\rm m}28.1^{\rm s}$ to $5^{\rm h}35^{\rm m}27.8^{\rm s}$ and declination with ticks on the vertical axis from $-69^\circ16^\prime12.0^{\prime\prime}$ to $-69^\circ16^\prime10.5^{\prime\prime}$.
(b) An image adapted from \cite{Matsuuraetal2024} of the F164N image in colors, revealing mainly the equatorial ring and the `Keyhole,' with ALMA 315~GHz continuum emission in contours. The double-sided arrows mark the main symmetry axis of SN 1987A and are located at the same place in both panels. }
\label{fig:SN1987ADust}
\end{figure}

Images of SN 1987A by ALMA at 315~GHz from  \cite{Bouchetetal2024} and \cite{Matsuuraetal2024} show emission, mainly from the equatorial ring around SN 1987A that was ejected twenty thousand years before the explosion. I present these images at the same scale in Figure \ref{fig:SN1987ADust}. In addition, these images reveal an elongated dust-ejecta structure closer to the center than the equatorial ring, which is aligned with the bipolar structure known as the `Keyhole,' seen at many other shorter wavelengths. Already Figure 2 of \cite{Ciganetal2019} showed the alignment of the ALMA 315~GHz dust emission with the H$\alpha$ emission of the `Keyhole.' The inner ejecta of SN 1987A, where the dust is, did not yet collide with any ambient gas, as it is inside the equatorial ring. Therefore, the dust formation is not due to ejecta collision with a circumstellar material. 

\cite{Matsuuraetal2024} discuss the presence of dust in the plane between the two `Keyhole' lobes of SN 1987A, what they term the bar (as marked on the lower panel of Figure \ref{fig:SN1987ADust}). It might be that the bar is a region compressed between the two lobes of the bipolar structure. According to the JJEM, a powerful pair of jets inflated the two lobes, compressing gas within them and between them, i.e., in the bar. 

Overall, I find that the dust in the ejecta of SN 1987A follows the morphology of the central bipolar structure, which the JJEM attributes to a powerful pair of jets that participated in the explosion of SN 1987A. 

\section{Summary} 
\label{sec:Summary}

In this \textit{Letter}, I focus on pointing out the point-symmetric, Figures \ref{Fig:CassiopeiaJWST} and \ref{fig:CrabDust}, and bipolar, Figure \ref{fig:SN1987ADust}, morphologies of dust in three CCSNRs. Accepting that jets shape these morphologies within the framework of the JJEM, this suggests that exploding jets enhance dust formation in CCSNRs. In this study, I do not address the entire chain of processes by which the jets enhance dust formation; I only argue that the jets compress the gas, and the high density facilitates dust formation in the jet-compressed zones. Future studies should examine relevant processes (e.g., \citealt{Cherchneffetal2025} for a recent study of dust formation in SNRs), such as the compression of ejecta by jets that inflate bubbles and the interaction between the bubbles. I limit this study to highlighting the jet-shaped morphology of dust distribution (not all dust in a given CCSNR, nor in all CCSNRs). 

The study of dust formation and survivability in CCSNe is a hot topic, with numerous new observations using ALMA and JWST (e.g., \citealt{Claytonetal2025, Medleretal2025, Sarangietal2025, Shahbandehetal2025, Szalaietal2025, Tinyanontetal2025}), and theoretical studies (e.g., \citealt{Kirchschlageretal2023, MartinezGonzalez2025, Zhaoetal2025}).   
CCSNe of a given type might differ by one to two orders of magnitude in the mass of dust they form (e.g., \citealt{Gomezetal2025}). I propose that the properties of the exploding jets and their interaction with the core and envelope (if it exists) are among the factors that determine the amount of dust formed and the dust morphology. 

This study contributes to the diversity of processes in which CCSN exploding jets are involved and to establishing the JJEM as the primary explosion mechanism of CCSNe. 

\section*{Acknowledgements}

I thank Ori Fox for raising the question of the dust mass produced by different CCSNe, Dima Shishkin for comments on the preprint,  and two referees for comments that improved the manuscript.  
I thank the Charles Wolfson Academic Chair at the Technion for the support.
\newline
This work is based in part on observations made with the NASA/ESA/CSA James Webb Space Telescope. The data were obtained from the Mikulski Archive for Space Telescopes at the Space Telescope Science Institute, which is operated by the Association of Universities for Research in Astronomy, Inc., under NASA contract NAS 5-03127 for JWST. These observations are associated with programs No. 1947 (Figure \ref{Fig:CassiopeiaJWST}),  No. 1714 (DOI:10.17909/6264-w578; Figure \ref{fig:CrabDust}), and  No. 1232 (DOI:10.17909/k6j3-vm72) and No. 1714 (DOI:10.17909/6264-w578) (Figure \ref{fig:SN1987ADust}).   


%
\bibliography{reference}{}
\bibliographystyle{aasjournal}
  


\end{document}